\documentclass[12pt,preprint]{aastex}

\shorttitle{Escape of Secondary CR Positrons}
\shortauthors{Kawanaka}

\usepackage{epsf}
\usepackage{epsfig}
\usepackage{amsmath}

\begin{document}

\title{Escape of Secondary Cosmic-Ray Positrons Produced in a Supernova Remnant}
\author{Norita Kawanaka\altaffilmark{1}}
\altaffiltext{1}{Racah Institute for Physics, The Hebrew University, Jerusalem, 91904, Israel} 
\email{norita@phys.huji.ac.il}

\begin{abstract}
We discuss the acceleration and escape of secondary particles, especially positrons produced by hadronic interactions in a supernova remnant (SNR) shock.  During the shock acceleration, protons would interact with ambient gas and produce charged secondary particles, which would also be accelerated in a SNR and injected into the interstellar medium as cosmic-rays (CRs).  Some previous studies showed that the resulting positron spectrum at the SNR shock is harder than the primary proton spectrum, and proposed that the positron excess observed by PAMELA can be explained with this process.  We calculate the energy spectra of CR protons and secondary CR positrons running away from the SNR into the interstellar medium according to the phenomenological model of energy-dependent CR escape.  We show that, on the contrary to the results presented previously, the observed spectra of secondary CR particles generated in SNRs would be softer than those of primary CR particles, which means that the rise in the positron fraction cannot be reproduced by this model.
\end{abstract}

\section{Introduction}
Recently the PAMELA collaboration reported that the cosmic-ray (CR) positron fraction $e^+/(e^+ + e^-)$ is rising with energy in the range of $\sim 7-100{\rm GeV}$ (Adriani et al. 2009), which cannot be reproduced when considering only the positron production during the propagation of CR protons in the Galaxy.  After that, Fermi Large Area Telescope (LAT) observed the electron spectrum and positron spectrum separately making use of Earth's magnetic field, and confirmed this trend in the energy range of 20-200GeV (Ackermann et al. 2012).  In addition, the cosmic-ray electrons plus positrons flux has been measured by Fermi/H.E.S.S./ATIC/PPB-BETS and they are also shown to have the excess in comparison with the standard CR propagation model (Abdo et al. 2009a; Ackermann et al. 2010; Aharonian et al. 2008,2009; Chang et al. 2008; Torii et al. 2008).  These observations likely suggest a new kind of source of CR electrons and positrons, such as nearby pulsars (Profumo 2008; Malyshev et al. 2009; Grasso et al. 2009; Hooper et al. 2009; Yuksel et al. 2009; Kawanaka et al. 2010; Heyl et al. 2010; Kisaka \& Kawanaka 2012), microquasars (Heinz \& Sunyaev 2002), gamma-ray bursts (Ioka 2010; Calvez \& Kusenko 2010), and dark matter annihilations/decays (see Bergstrom et al. 2008; Arkani-Hamed et al. 2009; Maede et al. 2010; Nardi et al. 2009; Grasso et al. 2009 and references therein).

Supernova remnants (SNRs) are also promising candidates of CR positron sources (Berezhko et al. 2003; Fujita et al. 2009; Shaviv et al. 2009; Hu et al. 2009; Blasi 2009; Biermann et al. 2009; Lee et al. 2011).  It is widely believed that Galactic CR nuclei and electrons are accelerated in SNRs by the diffusive shock acceleration (DSA) mechanism (Bell 1978; Blandford \& Ostriker 1978).  The theory of DSA (for reviews, see Blandford \& Eichler 1987; Malkov \& Drury 2001) can naturally derive the power-law spectrum of particles accelerated in the SNR shock.  The TeV gamma-ray detections from the shell of young SNRs by H.E.S.S. (Aharonian et al. 2004, 2005) and the GeV gamma-ray detections from middle-aged SNRs interacting with molecular clouds by Fermi \& AGILE (Abdo et al. 2009b, 2010a, 2010b, 2010c; Tavani et al. 2010; Giuliani et al. 2010) can be interpreted as the results of hadronic interactions between CR protons accelerated in SNRs and ambient protons.

Blasi (2009) has proposed the model of positron production which considers the hadronic interactions during the proton acceleration in the SNR shock.  In this model the acceleration of primary CR protons  and the production and acceleration of secondary positrons take place in the same region.  According to this study the resulting positron spectrum would be harder than the primary electron spectrum (locally having the same index as the primary protons) and therefore the positron fraction would rise with energy as the data reported by PAMELA.  Because this model can also predict the production of antiprotons and secondary nuclei (B, Ti etc.), it is suggested that  antiproton-to-proton ratio ($\bar{p}/p$), boron-to-carbon ratio (B/C) and titanium-to-iron ratio (Ti/Fe) would show the excess above $\sim 100{\rm GeV}$ per nucleon (Blasi \& Serpico 2009; Mertsch \& Sarkar 2009; Ahlers et al. 2009; see also Kachelriess et al. 2011, who argued that this process cannot reproduce the rising of the positron fraction using Monte Carlo simulations).  These predictions in CR spectral features would be tested by the ongoing observations by AMS-02, together with the more precise data of the positron fraction which will be also provided after its measurement.

In order to predict the spectrum of these CR components observed at the Earth, we should calculate their outgoing flux from SNRs which generally should vary with time according to the dynamical evolution of SNRs.  The previous studies shown above, however, only estimate the spectrum of CRs advected to the downstream of the SNR shock with totally time-independent picture.   In this study, we discuss the secondary CR positron production process inside the SNR taking into account the time-dependent escape of them as well as that of primary CR protons.  The escape of CR particles from SNRs and the observational feature expected from that have been investigated by several authors (Ptuskin \& Zirakashvili 2005; Reville et al. 2009; Ohira et al. 2010, 2011a, 2011b; Caprioli et al. 2010; Drury 2011; Kawanaka et al. 2011).  According to these studies, the particles accelerated in the shock would escape the shock in a time-dependent manner -- the higher energy particles can escape the SNR earlier.  As the shock slows down and the magnetic field decays, the lower energy particles can leave the shock gradually.  This process can make the CR spectrum in the interstellar medium softer than that inside the SNRs (Ohira et al. 2010; Caprioli et al. 2010), which can give the explanation for the observed CR spectral index being softer than that predicted from DSA theory ($\sim 2$) and the spectral breaks found in the GeV emissions from the middle-aged SNRs interacting with molecular clouds (Ohira et al. 2011).  Within this picture, secondary particles produced and accelerated in the SNR shock would also escape the source in a time-dependent manner.  Therefore, in order to predict the observed spectra of those secondary particles, we should calculate their time-dependent escape from SNRs, as well as their production and acceleration.

This paper is structured as follows: In Section 2, we describe the model of time-dependent escape of primary and secondary cosmic-ray particles from SNRs and show the transport equations of them.  In Section 3, we solve the transport equation for secondary cosmic-ray positrons, and predict their energy spectrum.  We discuss our results in Section 4, and Section 5 is devoted to the summary or this work.

\section{Model}
According to the standard theory of DSA, the scattering of particles is due to the turbulent magnetic field around the shock.  Especially, in the upstream region, the turbulence is considered to be excited by the accelerated particles themselves (Bell 1978).  Then, in the far upstream region which is sufficiently distant from the shock front, because the particle flux should be small, the turbulence would be so weak that the particles cannot be backscattered to the shock and escape the SNR.  In other words, in the CR escape scenario, we should impose the escape boundary in the upstream with a finite distance (hereafter, $l$) from the shock front.   Letting $D(p)$ be the diffusion coefficient   for a particle with momentum of $p$ and $u_{\rm sh}$ be the velocity of the shock front, the escape condition for a particle can be described as
\begin{eqnarray}
\frac{D(p)}{u_{\rm sh}}\gtrsim l,
\end{eqnarray}
where the left hand side is the characteristic diffusion length of a particle with momentum $p$.  The escape boundary $x=-l$ (here $x<0$ is the upstream region) is the point beyond which particle-scattering Alfv\'{e}n waves are absent and therefore the mean free path of particles is much larger than the size of a SNR.  Zirakashvili \& Ptuskin (2008) argued that the distance between the escape boundary and the shock front should be given roughly by the radius of the remnant $\sim R_{\rm sh}$.  In fact, because of the spherical geometry of a SNR shock, the particles which have diffused out from the shock to the distance comparable with the shock radius would not be able to cross the shock again and therefore escape the SNR into the interstellar medium (ISM).

\subsection{Escape of Primary Cosmic-Ray Protons}
First let us discuss the escape flux of cosmic-ray protons as well as the maximum attainable energy of them as functions of time following the procedure presented in Caprioli et al. (2009) and Ohira et al. (2010).  The transport equation for primary protons in the stationary case in the shock rest frame is
\begin{eqnarray}
u(x)\frac{\partial f_1}{\partial x}=\frac{\partial}{\partial x}\left[ D(p)\frac{\partial f_1}{\partial x} \right] +\frac{p}{3}\frac{du}{dx}\frac{\partial f_1}{\partial p}+Q_1,
\end{eqnarray}
where $f_1(x,p)$ is the distribution function of protons and $Q_1$ is the source term.  Here $u(x)$ is the fluid velocity.  The fluid velocity $u(x)$ is given by
\begin{eqnarray}
u(x)=\left\{
\begin{array}{ll}
u_1 & (x<0) \\
u_2 & (x>0), \\
\end{array} \right.
\end{eqnarray}
where $u_1$ and $u_2$ are constants.  Especially $u_1$ us equal to the speed of the shock front in the laboratory frame.  When considering the time-dependent escape scenario, we use the boundary condition $f_1(x=-l,p)=0$, and then the solution of this transport equation is given by
\begin{eqnarray}
f_1(x,p)=\frac{f_{1,0}(p)}{1-\exp (-u_1 l/D)} \left[ \exp \left( \frac{u_1 x}{D(p)} \right)- \exp \left(-\frac{u_1 l}{D(p)} \right) \right], \label{f_1}
\end{eqnarray}
where $f_{1,0}(p)$ is determined from the junction condition at $x=0$,
\begin{eqnarray}
f_{1,0}(p)=K(t)\exp \left[ -\gamma \int_{p_{\rm min}}^p \frac{d\ln p^{\prime}}{1-\exp \left[ -\frac{u_1 l}{D(p^{\prime})}\right]} \right] , \label{f10}
\end{eqnarray}
where $\gamma=3u_1/(u_1-u_2)$ and $K(t)$ is the time-dependent normalization factor which will be determined below\footnote{In the realistic situation, the minimum momentum of the integral appearing in Eq.(\ref{f10}) $p_{\rm min}$ should be determined by taking into account the particle injection process and should vary with time in general.  Because the variation of $p_{\rm min}$ can be regarded as the change of the overall normalization factor of the distribution function $f_{1,0}(p)$, in the following discussion we assume that $K(t)$ varies with time instead of $p_{\rm min}$ which is fixed as a constant.  This assumption does not affect the results presented below.}.  Here we can derive the escape flux at the boundary $x=-l$ as
\begin{eqnarray}
\phi_1(p)&=&u_1 \left. f_1\right|_{x=-l}-D(p)\left. \frac{\partial f_1}{\partial x}\right|_{x=-l} \nonumber \\
&=&-\frac{u_1 f_{1,0}(p)}{\exp (u_1l/D)-1},
\end{eqnarray}
and this value is negative which means that the flux is directed to the upstream (i.e. outward of the shock).  We assume the Bohm-type diffusion in which the mean free path of a particle is proportional to its Larmor radius.  Then the diffusion coefficient for particles with a momentum $p$ and a charge $e$ can be expressed as
\begin{eqnarray}
D(p)=\eta_g\frac{c^2 p}{3eB},
\end{eqnarray}
where $B$ is the magnetic field strength and $\eta_g$ is the gyrofactor which should be equal to unity for the Bohm limit.  In this case the absolute value of this escape flux $-\phi_1(p)$ would attain its maximum value when the momentum is
\begin{eqnarray}
p=\frac{u_1 l}{\gamma D_0}\equiv p_{1,m}, \label{primarymax}
\end{eqnarray}
where $D_0=D(p)/p$, and then the maximum value of the escape flux is
\begin{eqnarray}
-\phi_1(p_{1,m})=\frac{u_1 f_{1,0}(p_{1,m})}{e^{\gamma}-1},
\end{eqnarray}
 (Caprioli et al. 2009).  As $f_{1,0}(p)\propto p^{-\gamma}$ for $p\ll p_{1,m}$ and $f_{1,0}\propto \exp(-p/p_{1,m})$ for $p_{1,m} \ll p$, this quantity $p_{1,m}$ plays the role of maximum momentum of the primary accelerated particles, i.e. the accelerated protons with this momentum would escape the SNR into the ISM most efficiently.  According to Ohira et al. (2010), the distribution function of this escape flux can be approximated as
\begin{eqnarray}
-\phi_1(p)&\simeq &\frac{u_1f_{1,0}(p_{1,m})}{e^{\gamma}-1}\exp \left[ -\left( \frac{\ln p -\ln p_{1,m}}{\sigma} \right)^2 \right] \nonumber \\
&\simeq &\frac{u_1f_{1,0}(p_{1,m})}{e^{\gamma}-1}\cdot \sqrt{\pi}\sigma \cdot \delta(\ln p-\ln p_{1,m}), \label{lognormal}
\end{eqnarray}
where $\sigma^2=2(1-e^{-\gamma})/\gamma$.
 
 Because $D_0$ which appears in the expression of $p_{1,m}$ is proportional to $\eta_g B^{-1}$, the maximum momentum of escaping particles $p_{1,m}$ depends on the evolution of the amplification and decay of the magnetic field and turbulence around the shock (Ptuskin \& Zirakashivili 2003, 2005; Yan et al. 2012).  Here we assume that, during the Sedov phase, the maximum momentum of escaping primary particles decreases as $\propto t^{-\alpha}$ where $\alpha$ is a parameter which characterizes the evolution of $p_{1,m}$.  Note that, as is obvious from Eq. (\ref{primarymax}), this assumption is equivalent with the assumption that the combination of $u_1 l/D_0$ is proportional to $t^{-\alpha}$.  Now we consider the Sedov phase and assume that $l$ is proportional to the shock radius $R_{\rm sh}$, the coefficient $D_0$ evolves with time being proportional to $t^{\alpha-1/5}$.  This time dependence can be interpreted also in the following way; the maximum momentum of escaping particles should be determined by the condition that the timescale for a particle to escape the source with size of $\sim R_{\rm sh}$, $t_{\rm esc}\sim R_{\rm sh}^2/D(p)$, is comparable with the acceleration timescale in the SNR shock, $t_{\rm acc}\sim D(p)/u_{\rm sh}$.  Considering the time dependence of $R_{\rm sh}$ and $u_{\rm sh}$ in the Sedov phase, we have the time dependence of the maximum momentum which can escape the source as $p_{\rm max}\propto D_0^{-1}t^{-1/5}$.  Therefore, when imposing that $p_{\rm max}\propto t^{-\alpha}$, $D_0$ should be proportional to $t^{\alpha-1/5}$.
 
 On the other hand, the normalization factor of the escape flux would evolve with time approximately as $\propto u_1 K(t)p_{1,m}^{-\gamma}$.  Considering that the energy spectrum of escaping particles is described as
 \begin{eqnarray}
 \frac{dN_{{\rm esc},1}}{d\varepsilon dt}\propto 4\pi R_{\rm sh}^2 \cdot 4\pi p^2 \phi_1(p), \label{luminosity}
 \end{eqnarray}
where $\varepsilon=\sqrt{(cp)^2+(m_p c^2)^2}$ is the energy of a particle, and that $R_{\rm sh}$ and $u_{\rm sh}$ evolve as $\propto t^{2/5}$ and $\propto t^{-3/5}$ in the Sedov phase, respectively, we can derive the power-law index of the time-integrated energy spectrum of primary cosmic-ray protons $s_{{\rm esc},1}$ as
 \begin{eqnarray}
 s_{{\rm esc},1}\simeq (\gamma-2)+\frac{1}{\alpha} \left(\tilde{\beta}+\frac{6}{5} \right), \label{primaryindex}
 \end{eqnarray}
 where we use Eq. (\ref{lognormal}) and assume that $K(t)$ is proportional to $t^{\tilde{\beta}}$.  If we redefine as $s\equiv\gamma-2$ and $\beta\equiv \tilde{\beta}+6/5$, this formula is identical to Eq. (28) in Ohira et al. (2010).  In the standard theory of non-relativistic DSA, the index $\gamma$ should be equal to 4 and therefore the energy spectral index of accelerated particles should be 2.  However, if the normalization factor of CR particle flux increases with time (i.e. if $\beta>0$), the spectral index of escaping CR particles would be softer than 2.{\footnote In some young SNRs, the synchrotron emission spectra observed from inside their shells have a steeper spectral indices than 0.5, which is expected from electrons with a spectral index 2.  As those SNRs are still in the pre-Sedov phase or very early stage of the Sedov phase, the energy-dependent escape scenario is not suitable to account for those observations.  Some other models have been proposed to explain them (Kirk et al. 1996; Zirakashvili \& Ptuskin 2009; Bell et al. 2011; Ohira 2012).}  We cannot tell the relevant value of $\beta$ from observations or theories because the particle injection processes for the Fermi acceleration have not been well understood.  Here we consider the thermal leakage model of the particle injection (Malkov \& V\"olk 1995).  In this model, it is required that the distribution function of accelerated particles is continuous to the downstream Maxwell distribution at the injection momentum, which is assumed to be proportional to the shock velocity, which decreases with time as $\propto t^{-3/5}$ during the Sedov phase.  We assume $\beta=0.6$ (i.e. $\tilde{\beta}=-0.6$) during the Sedov phase, which is allowed in the thermal leakage model (Ohira et al. 2010).  Adopting this value with $\gamma=4$ and $\alpha=2.6$ (see Sec. 3), the spectral index of primary protons is $s_{{\rm esc}.1}\simeq 2.23$, which is consistent with that at the source expected from observations (Ohira et al. 2010).

\subsection{Transport Equation of Secondary Cosmic-Ray Particles}
During the acceleration of protons in the SNR shocks, secondary particles such as positrons, antiprotons are produced through hadronic interactions.  In order to calculate their spectrum, we should solve the transport equation with a source term which is related to the primary proton distribution function derived above:
\begin{eqnarray}
u(x)\frac{\partial f_2}{\partial x}=\frac{\partial}{\partial x}\left[ D(p)\frac{\partial f_2}{\partial x} \right] +\frac{p}{3}\frac{du}{dx}\frac{\partial f_2}{\partial p}+Q_2(x,p). \label{2ndary}
\end{eqnarray}
Here $f_2(x,p)$ is the distribution function of secondary particles and $Q_2(x,p)$ is the source term which can be determined from Eq.(\ref{f_1}) as
\begin{eqnarray}
Q_2(x,p)=\left \{
\begin{array}{ll}
Q_{2,u}(p) \frac{\exp \left[ \frac{u_1x}{D(p_{\rm p})} \right] -\exp \left[ -\frac{u_1l}{D(p_{\rm p})} \right] }{1-\exp \left[ -\frac{u_1 l}{D(p_{\rm p})}\right] } & (x<0) \\
Q_{2,d}(p) & (x>0), \\
\end{array} \right.
\end{eqnarray}
where $p_{\rm p}$ is the momentum of a secondary particle generated from a primary particle with a momentum $p$.  Generally $p$ and $p_{\rm p}$ can be approximately related linearly: $p \approx \xi_i p_{\rm p}$.  In the case of positron production, $\xi_{e^+} \approx 0.05$,  and in the case of $\bar{p}$, $\xi_{\bar{p}} \approx 0.17$.  Here we assume $D(p)\propto p$ (Bohm diffusion), so we can replace $D(p_{\rm p})$ by $D(p)/\xi_i$.  The momentum dependences $Q_{2,u}(p)$ and $Q_{2,d}(p)$ can be expressed as
\begin{eqnarray}
Q_{2,u}(p)&=& \frac{c n_{{\rm gas},u}}{4\pi p^2}\int _{p}^{\infty} dp^{\prime} N_1(p^{\prime} ) \frac{d\sigma_{pp\rightarrow i}}{dp} \label{secondaryprod} \\
Q_{2,d}(p)&=& r Q_{2,u}(p),
\end{eqnarray}
where $N_1(p)=4\pi p^2 f_1(0,p)$ is the number density of primary CR particles per unit momentum, $d\sigma_{pp}/dp$ is the differential cross section of $i$ particle generations via $pp$ interactions, and $r$ is the compression ratio of the shock, respectively.

The solution of Eq.(\ref{2ndary}) should satisfy the following boundary conditions:
\begin{eqnarray}
&{\rm (i)}& \lim_{x \to -0} f_{2,u} = \lim_{x \to +0} f_{2,d}, \nonumber \\
&{\rm (ii)}& \left[ D(p)\frac{\partial f_2}{\partial x} \right]_{x=+0}^{x=-0}=\frac{1}{3}(u_2-u_1)p \left. \frac{\partial f_2}{\partial p} \right| _{x=0}, \nonumber \\
&{\rm (iii)}& \lim_{x \to -l} f_{2,u}=0, \nonumber \\
&{\rm (iv)}& \left| \lim_{x \to \infty} f_2 \right | < \infty. \nonumber
\end{eqnarray}
Here the condition (ii) comes from the integration of Eq.(\ref{2ndary}) across the shock front, and it yields the differential equation with respect of $p$.

\section{Energy Spectrum of Escaping Secondary Positrons}
Under the boundary conditions shown above, we can solve the transport equation analytically as
\begin{eqnarray}
f_{2,u}&=&Y_1(p)\exp\left[ \frac{u_1 x}{D}\right] +Y_2(p) \exp\left[ \frac{\xi_i u_1 x}{D}\right]  +Y_3(p) x+Y_4(p), \\
f_{2,d}&=&f_{2}(x=0,p)+\frac{Q_{2,d}(p)}{u_2}x,
\end{eqnarray}
where
\begin{eqnarray}
Y_1(p)&=&\frac{1}{1-\exp(-u_1 l/D)} \left[f_2(0,p)-\frac{D}{u_1^2}\frac{Q_{2,u}(p)}{\xi_i(1-\xi_i)} +\frac{l}{u_1}\frac{Q_{2,u}(p)}{\exp(\xi_iu_1 l/D)-1}\right], \\
Y_2(p)&=&\frac{D}{u_1^2}\frac{Q_{2,u}(p)}{1-\exp(-\xi_iu_1 l/D)}\frac{1}{\xi_i(1-\xi_i)}, \\
Y_3(p)&=&-\frac{1}{u_1}\frac{Q_{2,u}(p)}{\exp(\xi_iu_1 l/D)-1}, \\
Y_4(p)&=&f_2(0,p)\left(1-\frac{1}{1-\exp(-u_1 l/D)}\right) \nonumber \\
&&+\frac{1}{1-\exp (-u_1 l/D)}\left(\frac{D}{u_1^2}\frac{Q_{2,u}(p)}{\xi_i-\xi_i^2} -\frac{l}{u_1}\frac{Q_{2,u}(p)}{\exp (\xi_i u_1 l/D)-1}\right) \nonumber \\
&&-\frac{D}{u_1^2}\frac{Q_{2,u}(p)}{1-\exp (-\xi_i u_1 l/D)}\frac{1}{\xi_i-\xi_i^2}, \\
f_2(0,p)&=&-\gamma \int_{p_{\rm min}}^p \frac{dp^{\prime}}{p^{\prime}}\frac{H(p)}{H(p^{\prime})}\frac{Q_{2,u}(p^{\prime})G(p^{\prime})}{u_1}, \label{f_20} \\
G(p)&=&\frac{1}{1-\exp (-u_1 l/D)}\left[ -\frac{D}{u_1}\frac{1}{\xi_i -\xi_i^2}+\frac{l}{\exp(\xi_i u_1 l/D)-1}\right] \nonumber \\
&&+\frac{D}{u_1}\left[ 1+\frac{\xi_i}{1-\xi_i}\frac{1}{1-\exp (-\xi_i u_1 l/D)}\right] -\frac{r^2 D}{u_1}, \\
H(p)&=&\exp \left[ -\gamma \int_{p_{\rm min}}^p \frac{dp^{\prime}}{p^{\prime}} \frac{1}{1-\exp (-u_1 l/D)} \right],
\end{eqnarray}
and so the escape flux of the secondary particles at $x=-l$ is
\begin{eqnarray}
\phi_2(p)&=&u_1 f_{2,u}|_{x=-l}-D(p)\left. \frac{\partial f_{2,u}}{\partial x} \right|_{x=-l} \nonumber \\
&=&-\frac{u_1}{\exp (u_1 l/D)-1}\left[ f_2(0,p)-\frac{D}{u_1^2}\frac{Q_{2,u}}{\xi_i -\xi_i^2}+\frac{l}{u_1}\frac{Q_{2,u}}{\exp(\xi_i u_1 l/D)-1} \right] \nonumber \\
&&-\frac{\xi_i D}{u_1(1-\xi_i)}\frac{Q_{2,u}}{\exp (\xi_i u_1 l/D)-1}, \label{secondaryflux}
\end{eqnarray}
and in order to derive the spectrum of the CR particles injected from a SNR into the interstellar medium, we should integrate this escaping CR flux with time.  

In the following calculations, we assume the explosion energy of a supernova as $E_{\rm SN}=10^{51}{\rm erg}$, the number density of the interstellar matter as $n=1{\rm cm}^{-3}$, and the ejecta mass as $M_{\rm ej}=1M_{\odot}$.  With this environmental parameters, the SNR shock radius at the beginning of the Sedov phase, $R_{\rm S}$, and the SNR age at that time $t_{\rm S}$ can be expressed as
\begin{eqnarray}
R_{\rm S}&=&2.13{\rm pc}~\left(\frac{M_{\rm ej}}{1M_{\odot}}\right)^{1/3}\left(\frac{n}{1{\rm cm}^{-3}}\right)^{-1/3}, \\
t_{\rm S}&=&209{\rm yr}~\left(\frac{E_{\rm SN}}{10^{51}{\rm erg}}\right)^{-1/2} \left( \frac{M_{\rm ej}}{1M_{\odot}} \right)^{5/6}\left( \frac{n}{1{\rm cm}^{-3}} \right)^{-1/3}. \label{sedovtime}
\end{eqnarray}

In addition, we perform the calculations assuming $\alpha=2.6$ and
\begin{eqnarray}
cp_{1,m}=10^{6.5}{\rm GeV}\times \left(\frac{t}{t_{\rm S}} \right)^{-2.6},
\end{eqnarray}
which is also adopted in Gabici et al. (2009) and Ohira et al. (2010) according to the hypothesis that SNRs are responsible for the observed CRs with energies from $\sim 1{\rm GeV}$ up to the knee energy $(\sim 10^{15.5}{\rm GeV})$.  In this case, the lowest energy CR particles ($\varepsilon\sim 1{\rm GeV}$) would escape the SNR at the end of the Sedov phase $t=10^{2.5}t_{\rm S}$, when the shock radius $R_{\rm sh}$ becomes $10R_{\rm S}$.  Note that the magnetic field strength at the beginning of the Sedov phase should be amplified up to
\begin{eqnarray}
B_{\rm S}\simeq 174\mu{\rm G}~\left( \frac{\eta_g(t_{\rm S})}{1} \right) \left( \frac{cp_{1,{\rm max}}(t_{\rm S})}{10^{6.5}{\rm GeV}}\right)\left(\frac{t_{\rm S}}{200{\rm yr}}\right) \left( \frac{R_{\rm S}}{2{\rm pc}}\right)^{-2}\left(\frac{l}{0.1R_{\rm S}}\right)^{-1}.
\end{eqnarray}
In calculating the positron production from hadronic interactions in the SNR by Eq.(\ref{secondaryprod}), we used the parametrization for the cross section provided by Kamae et al. (2006).  Also, we assume that $r=4$ and $\gamma=4$, which is true in the case of non-relativistic strong shock, which is realized in a typical SNR.

\subsection{Escape Flux}
In Figure 1, we show the escape fluxes of primary CR protons and secondary CR positrons as functions of momentum.  In the latter one, we can see two peaks at different momentums.  These peaks come from two dominant terms in Eq.(\ref{secondaryflux}):
\begin{eqnarray}
-\phi_{2,A}(p)&\equiv&\frac{u_1 f_2(0,p)}{\exp(u_1l/D)-1}, \\
-\phi_{2,B}(p)&\equiv&\frac{\xi_{e^+}D}{u_1(1-\xi_{e^+})}\frac{Q_{2,u}(p)}{\exp(\xi_{e^+}u_1l/D)-1},
\end{eqnarray}
where $\phi_{2,A}(p)$ contributes to the peak at a higher momentum and $\phi_{2,B}(p)$ to the peak at a lower momentum.  Judging from their functional forms and the figure, $\phi_{2,A}$ has a similar form to the escape flux of primary protons $\phi_1$, and so this component can be regarded as the positrons accelerated at the shock in the same manner as protons.  On the other hand, $\phi_{2,B}$ has a peak with momentum of $\sim 0.05p_{1,{\rm max}}\sim \xi_{e^+}p_{1,{\rm max}}$.  Therefore, we can say that the positrons in this component has been produced mainly by protons with momentum around $p_{1,{\rm max}}$.  Note that, though in Fig. 1 the peak of $\phi_{2,B}$ seems to be higher than that of $\phi_{2,B}$, the escape flux per unit energy bin is proportional to $p^2\times \phi_2(p)$, and then the main contribution to the resulting energy spectrum of escaping positrons comes from $\phi_{2,A}$, not from $\phi_{2,B}$ (see below). 

In Figure 2 we show the energy distribution functions of the escape flux ($\propto p^2 \phi(p)$) of primary protons (upper panel) and secondary positrons (lower panel) at $t=10t_{\rm S}$, $30t_{\rm S}$, $100t_{\rm S}$, and $300t_{\rm S}$.  In these calculations, we neglect the energy loss of positrons during acceleration due to synchrotron radiation.  If we take into account this effect, positrons can escape the SNR only when their escape timescale is shorter than their cooling timescale:
\begin{eqnarray}
t_{\rm esc}\sim \frac{R_{\rm sh}^2}{D(p)}\lesssim t_{\rm cool}\sim \frac{9m_e^4 c^7}{4e^4 B_{\rm d}^2 \varepsilon_{e^+}},
\end{eqnarray}
where $m_e$ and $\varepsilon_{e^+}$ are the mass and the energy of a positron, respectively, and $B_d=4B$ is the magnetic field strength in the downstream region.  Ohira et al. (2011) showed that the highest energy of electrons which can escape the SNR is $E_{\rm m,e}\sim 50{\rm TeV}$.  Therefore, we cannot observe the positrons which are more energetic than this $E_{\rm m,e}$.  As long as discussing positrons with energy lower than this value, the results shown here describe the phenomena well.

We can see that, (1) the energy distribution function of the secondary CR positron flux has a peak at a certain energy in each time step, and that (2) the peak energy decreases with time in the same rate as that of primary protons.  The fact (1) means that, as in the case of primary CR protons, there is the maximum momentum also for the secondary CR positrons produced in the SNR.  As for (2), it can be proved in the following way.  First we can approximate the source function for secondary positrons $Q_{2,u}(p)$ as
\begin{eqnarray}
Q_{2,u}(p)\simeq \mathcal{R}cn_{\rm gas}\sigma_{pp}K(t)\exp \left[ -\gamma \int_{p_{\rm inj}}^p \frac{d\ln p^{\prime}}{1-\exp (-\xi_{e^+}u_1 l/D)} \right], \label{secondaryapprox}
\end{eqnarray}
where we assume that the spectrum of secondary positrons has the same form as that of primary particles except for an energy-shift factor of $\xi_{e^+}$.  Here $\mathcal{R}$ is a dimensionless constant which should be determined by the microphysics.  Then, from Eq. (\ref{secondaryflux}), the escape flux of secondary CR positrons $\phi_2(p)$ can be written as
\begin{eqnarray}
\phi_2(p)=Z(t)\cdot \Phi(D_0 p/u_1 l)+C(t),
\end{eqnarray}
where $Z(t)$ and $C(t)$ are functions of time $t$ and do not depend on $p$, and $\Phi(y)$ is a function which does not depend on time.  In other words, we can rewrite $\phi_2(p)$ so that the momentum $p$ appears only in the combination of $D_0 p/u_1 l$.  This fact shows that the peak momentum of $\phi_2(p)$ (i.e. the momentum at which $d\phi_2(p)/dp=0$) is given by $p_{2,m}=y_mu_1l/D_0$, where $y_m$ is the value of $y$ at which $\Phi(y)$ attains its peak value.  Comparing this with Eq.(\ref{primarymax}), we can say that the maximum momentum of secondary positrons decreases in the same way as that of primary protons, $p_{2,m}\propto t^{-\alpha}$.

\subsection{Time Integrated Spectrum of Escaping Positrons}
Now we can numerically calculate the escape flux of secondary positrons and its time evolution using the equations presented above.  In Figure 3 we show the time integrated spectra of escaping protons and positrons.  We can see that the spectral index of escaping secondary positrons is {\it softer} than that of escaping protons.  Considering that the spectral index of primary electrons would be the same as primary protons, this fact shows that the process discussed here cannot make the rise in the positron fraction ($n_{e^+}/(n_{e^-}+n_{e^+})$) with energy.  This result is on the contrary to those presented in Blasi (2009) and Ahlers et al. (2009), in which they argued that the positron production by hadronic interactions during the acceleration of primary protons inside the SNR can reproduce the positron excess observed by PAMELA.

This result can be interpreted in the following way.  As shown in Fig. 1, the peak of the escaping positron flux is dominated by the first term of Eq.(\ref{secondaryflux}).  Using the approximation Eq. (\ref{secondaryapprox}), we can see that $f_{2,0}(p)\propto p^{-\gamma+1}$ for $p\ll u_1l/D_0$, while $f_{2,0}(p)\propto \exp(-\gamma D_0 p/u_1 l)$ for $p\gg u_1l/D_0$.  Especially, at the peak momentum, the normalization factor of $\phi_2$ would evolve with time approximately as $\propto u_1\cdot (D_0/u_1^2)K(t)p_{2,m}^{-\gamma+1}$.  Considering that the spectrum of escaping particles should be determined by integrating Eq. (\ref{luminosity}) with time, and that $R_{\rm sh}\propto t^{2/5}$, $u_1\propto t^{-3/5}$, $p_{2,m}\propto t^{-\alpha}$, and $D_0\propto t^{\alpha-1/5}$, we can estimate the power-law index of the time-integrated energy spectrum of escaping positrons $s_{{\rm esc},2}$ in the similar way as that of primary protons:
\begin{eqnarray}
s_{{\rm esc},2}&\simeq &(\gamma-1-2)+\frac{1}{\alpha} \left( \tilde{\beta}+\alpha+\frac{11}{5}\right) \nonumber \\
&=&(\gamma-2)+\frac{1}{\alpha}\left( \tilde{\beta}+\frac{11}{5} \right),
\end{eqnarray}
which is obviously softer than that of escaping primary protons shown in (\ref{primaryindex}).  This additional softening, $s_{{\rm esc},2}-s_{{\rm esc},1}=1/\alpha$, is independent of the value of $\tilde{\beta}$, and this comes from the additional factor $D(p_{2,m})/u_1^2$ to their escape flux which increases with time as $t^1$.  Because the momentum of secondary particles is roughly proportional to that of their parent particles, this factor is proportional to the acceleration timescale of protons which produce the positrons with momentum of $p_{2,m}$.  Especially, this timescale is comparable with the escape timescale, $t_{\rm esc}\sim R_{\rm sh}^2/D$.  Roughly speaking, the amount of secondary CR particles produced in the SNR is proportional to the time during which primary CR particles interact with the ambient gas in the SNR.  For lower energy primary protons, they would stay in the SNR longer (i.e. their escape timescale is longer), and so they can produce more secondary positrons than higher energy protons, which would stay in the SNR only for a shorter time and escape it into the ISM earlier.  As a result, the normalization factor of the escaping positron flux increases faster than that of the primary proton flux, and therefore the time-integrated positron spectrum would be softer than the proton spectrum.  The difference between the spectral indices of primary protons and secondary positrons is determined by the index $\alpha$, which characterizes the evolution of the maximum momentum of escaping CR particles.  Physically, we can say that the slower the decay of magnetic field and magnetohydrodynamic turbulence
 around the shock which determines the evolution of the diffusion coefficient there is, the softer the spectrum of escaping secondary particles would be.

\subsection{Effects of Particles Advected Downstream}
In the discussion above, we have only argued the spectrum of CR particles escaping the SNR right after accelerated at the shock.  Among the accelerated CR particles, those with the momentum lower than $p_{\rm max}(t)$ are confined in the SNR and lose their energy adiabatically (Ptuskin \& Zirakashvili 2005).  Strictly speaking, in the later time some of them satisfy the condition for the escape, $p>p_{\rm max}(t)$, and then they can escape the SNR into the ISM together with those accelerated up to $p_{\rm max}(t)$ at that time.  Here we estimate the contribution of such CR component and compare it with that of CR particles escaping the SNR right after the acceleration.

As shown in the previous section, the distribution function of accelerated secondary particles at the shock can be approximated as
\begin{eqnarray}
f_{2,0}(p,t)\simeq At^{\alpha+\tilde{\beta}+1}p^{-\gamma+1}, \label{f20app}
\end{eqnarray}
where $A$ is a normalization factor which is constant with time.  Here we use the facts that in the Sedov phase the factors that appear in Eq. (\ref{f20app}) evolve as $u_{\rm sh}\propto t^{-3/5}$, $D_0\propto t^{\alpha-1/5}$, and $K(t)\propto t^{\tilde{\beta}}$.  

On the other hand, the particles advected downstream lose their energy adiabatically as
\begin{eqnarray}
\frac{dp}{dt}=-\frac{\nabla \cdot \mbox{\boldmath $u$}}{3}p,
\end{eqnarray}
where $\mbox{\boldmath $u$}$ is the fluid velocity inside the SNR shell.  Here we consider the case of a spherically symmetric SNR shock and so $\mbox{\boldmath $u$}$ has only the radial component:
\begin{eqnarray}
u(R)=\left(1-\frac{1}{r}\right) \frac{u_{\rm sh}(t)}{R_{\rm sh}(t)}R,
\end{eqnarray}
where $R$ is the radial coordinate in the SNR.  This gives us the energy of the particles around the shock $R=R_{\rm sh}$ as a function of time:
\begin{eqnarray}
p_{\rm sh}(t)=p_{\rm sh}(t_S)\left( \frac{t}{t_S} \right)^{-2(1-1/r)/5}.
\end{eqnarray}
Using these results, we can evaluate the distribution function of particles that are accelerated at $t^{\prime}<t$ and advected downstream, $f_{\rm SNR}(p,t;t^{\prime})$, at the shock front.   As the total number of CR particles inside the SNR is conserved,
\begin{eqnarray}
f_{\rm SNR}(p,t;t^{\prime})R_{\rm sh}^2 dR_{\rm sh}p^2 dp=f_{2,0}(p^{\prime},t^{\prime}){R_{\rm sh}^{\prime}}^2 dR_{\rm sh}^{\prime} {p^{\prime}}^2 dp^{\prime},
\end{eqnarray}
where $R_{\rm sh}^{\prime}=R_{\rm sh}(t^{\prime})$ and $p^{\prime}=p(t^{\prime}/t)^{-2(1-1/r)/5}$ is the momentum of particles around the shock that had the momentum of $p$ at the time of $t^{\prime}$.  Then we have
\begin{eqnarray}
f_{\rm SNR}(p,t;t^{\prime})&=&f_{2,0}(p^{\prime},t^{\prime})\left( \frac{t^{\prime}}{t} \right)^{6/5r} \nonumber \\
&=&A{t^{\prime}}^{\alpha+\tilde{\beta}+1}p^{-\gamma+1} \left( \frac{t^{\prime}}{t} \right)^{\frac{2}{5}(\gamma-1)(1-1/r)+\frac{6}{5r}},
\end{eqnarray}
and this would be larger than $f_{2,0}(p,t)$ when
\begin{eqnarray}
\tilde{\beta}>-\alpha-1-\frac{2}{5}(\gamma-1)\left(1-\frac{1}{r} \right)-\frac{6}{5r},
\end{eqnarray}
or, using $\beta=\tilde{\beta}+6/5$ and $\gamma=4$ for a non-relativistic shock, we can simply write the condition as
\begin{eqnarray}
\beta>-\alpha-1.
\end{eqnarray}
The right-hand side of this inequality is negative because we are assuming the positive $\alpha$ (i.e. $p_{\rm max}(t)$ decreases with time).  Therefore, as long as we consider the case with positive $\beta$, which is required to account for the observed cosmic-ray spectrum in the context of the energy-dependent CR escape scenario, we can conclude that the flux of CR particles escaping right after the acceleration always always dominates the flux of those advected downstream and suffering from adiabatic cooling inside the SNR.  

\section{Discussion}
The results presented above show that the positrons which are produced during the acceleration of protons inside the SNR and emitted into the ISM would have a softer energy spectrum than those of primary protons and electrons.  As is obvious from the discussions above, the same logic can be applied to any secondary particles which are also produced during the acceleration of protons in the SNRs.  Hence, if the CR particles accelerated in the SNR shock escape the shock in an energy-dependent way, the emergent secondary electron plus positron cannot reproduce the `excess' in the electron spectrum reported by Fermi LAT (Abdo et al. 2009a; Ackermann et al. 2010) because they should have a softer spectrum than primary electrons.  In addition, the spectrum of antiprotons produced during the acceleration of protons and emitted from SNRs would be softer than primary protons, and so the ratio between them $\bar{p}/p$ would decreases with energy, which is on the contrary to the results presented in Blasi \& Serpico (2009).

Our results are partly consistent with those presented in previous studies such as Blasi (2009).  In fact, in our formalism, the momentum distribution function of positrons {\it at the SNR shock} is expressed as Eq.(\ref{f_20}), and we can see that in the limit of $l\rightarrow \infty$ this is consistent with Eq.(6) in Blasi (2009).  As this function is approximately proportional to $\sim p^{-\gamma+1}$ when $D(p)$ is linearly dependent on $p$, they conclude that the secondary positron (and electron) spectrum would be harder than the primary electron spectrum.  However, this distribution function is generally different from that in the ISM.  For example, taking into account the energy-dependent escape of CR particles from a SNR, even the spectrum of primary CR particles would be softer than that in the source because more and more particles are injected into the acceleration process as the shock sweeps the ISM (Ohira et al. 2010).  In the case of secondary CR particles produced in a SNR, not only the increasing swept-up ISM gas particles injected into the shock, but also the evolution of the factor $D_0/u_{\rm sh}^2$ in the distribution function of secondary particles, which does not appear in that of primary particles, is also responsible for the softening of the energy spectrum of CR particles in the ISM.  Considering the deceleration of the SNR shock and the decay of the magnetic field and turbulence around the shock, it is quite natural to assume that this factor increases with time, and this is just what has never been taken into account in the previous studies on this secondary CR production process in a SNR.  As is mentioned in the previous section, because the lower energy positrons escape the SNR later, this factor should be larger and then the number flux of escaping positrons would be larger, which makes their time-integrated energy spectrum softer.  In the calculations we simply assume that the escape momentum of CRs decreases with time as $t^{-\alpha}$ and the normalization factor of their escape flux increases as $t^{\tilde{\beta}}$.  Although this CR escape model is very simplified and phenomenological, it can account for the gamma-ray observations of middle-age SNRs interacting with molecular clouds (Gabici et al. 2009; Ohira et al. 2011), and it is worthwhile to discuss this scenario in the various context.  Moreover, even if we do not assume the simple power-law dependence on time of the escape momentum, the discussions presented above still hold good as long as the CR particles escape the acceleration site in order from high energies to low energies.

We should mention the total energy of CR positrons emitted from SNRs.  As the spectral indices of CR protons and positrons are softer than 2, most of the energy of these CR particles are carried by those with the lowest energy $\varepsilon_{\rm min}$, and therefore their total energy can be estimated as $E_{\rm CR}\sim \varepsilon_{\rm min}^2dN/d\varepsilon|_{\varepsilon_{\rm min}}$.  From Fig. 3, where we calculate the spectra with typical environmental parameters (see Sec. 3), we can say that the ratio between the total energy of CR positrons escaping the SNR to that of secondary CR protons is $\lesssim 0.06\%$ in our case, which is a significant fraction of observed CR positron energy ratio to protons, $\sim 0.1\%$.  Considering the positron production process presented here, this ratio should be proportional to the product of the number density of the ISM and the escape time of the lowest energy particles, which is equal to the end of the Sedov phase: $E_{{\rm CR},e^+}/E_{{\rm CR},p}\propto nct_{\rm esc}(\varepsilon_{\rm min})\sim nc\times 10^{2.5}t_{\rm S}$.  From Eq.(\ref{sedovtime}), we can derive the dependence of the amount of positrons on the environmental parameters of a SNR as $\propto E_{\rm SN}^{-1/2}n^{2/3}$; if the supernova is less energetic or if the number density of the ISM around the supernova is larger, the amount of secondary positrons emitted from the SNR becomes larger.

Our discussion may be applied also to the acceleration and escape of secondary CR nuclei such as boron nuclei (B) which can be produced by spallation of carbon nuclei (C) inside the SNR (Mertsch \& Sarkar 2009).  It may be worthwhile to calculate the total number and energy spectrum of CR borons emitted from SNRs within this scenario and give the theoretical lower limit to the B/C ratio in the CR observed at the Earth, which will be explored in the wider energy range by AMS-02.  This is beyond the scope of this paper and left to the future work.

\section{Summary}
In this paper we have investigated both the acceleration and energy-dependent escape of secondary positrons produced in the SNR for the first time.  Especially we show that, if we assume that the CR particles with higher energy would escape earlier than lower energy particles into the ISM, the energy spectrum of secondary-produced particles escaping the SNR would be inevitably softer than that of primary protons.  This implies that the rising of the positron fraction in the energy range of $\sim 7-200{\rm GeV}$ measured by PAMELA and Fermi LAT cannot be accounted by the positron production during the acceleration of CR protons inside the SNR.  Therefore, if this CR escape scenario is true, it is implied that in order to account for this anomaly we need other primary sources of positrons with a hard spectral index, such as pulsars, dark matter particles, and so on.  With the similar discussion, we can conclude that the CR electron plus positron spectrum cannot be hard enough to reproduce the results of Fermi LAT, and that the antiproton ratio $\bar{p}/p$ does not rise with energy within this scenario.  These results are on the contrary to those presented in previous studies which have investigated this process.

NK is grateful to K. Ioka, Y. Ohira, and T. Piran for useful discussions and their precious comments.  This work is partially supported by an ERC advanced research grant.

\clearpage
\begin{figure}
 \begin{center}
  \includegraphics[width=160mm]{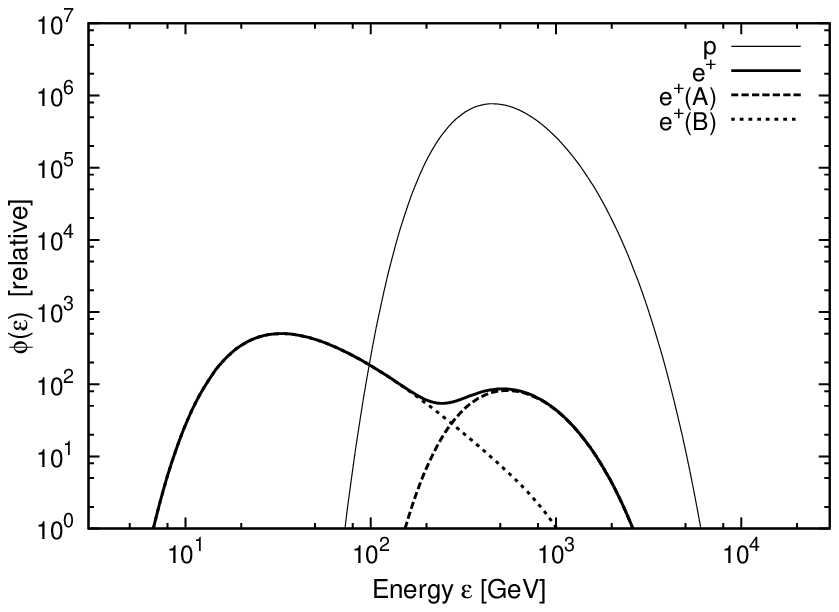}
 \end{center}
 \caption{Escape flux of primary CR protons $\phi_1(p)$ (thin solid line) and secondary CR positrons $\phi_2(p)$ (thick solid line) at the time $t=30t_{\rm S}$.  The dominant components which make double peaks in the positron flux are also shown; $\phi_{2,A}(p)$ (dashed line) and $\phi_{2,B}(p)$ (dotted line).}
 \label{fig1}
\end{figure}

\clearpage
\begin{figure}
\begin{center}
\includegraphics[width=250mm]{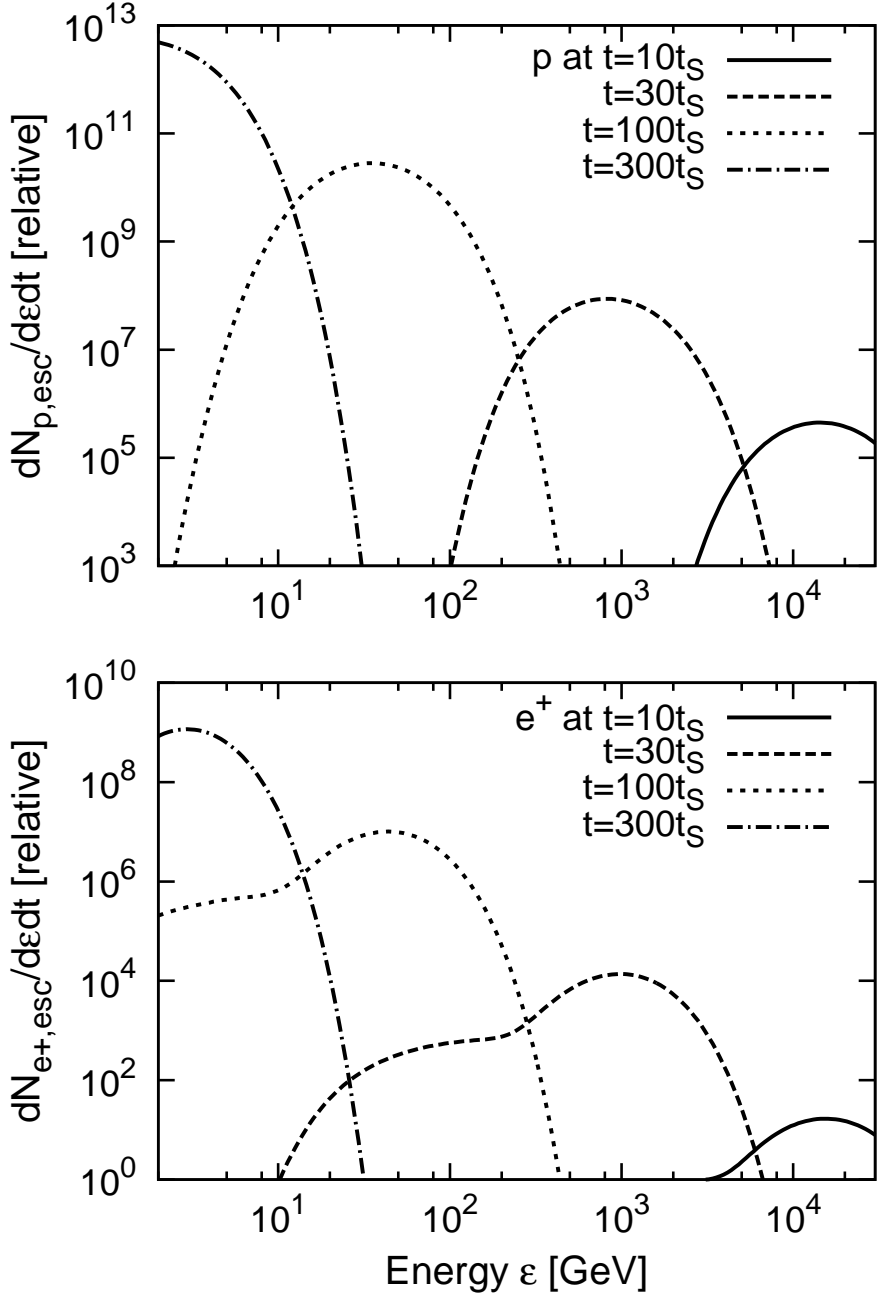}
\end{center}
\caption{Energy distribution functions of escaping CR protons (upper panel) and positrons (lower panel) emitted from a SNR per unit time at $t=10t_{\rm S}$ (solid lines), $30t_{\rm S}$ (dashed lines), $100t_{\rm S}$ (dotted lines) and $300t_{\rm S}$ (dot-dashed lines).  Here we assume $\beta=0.6$ (i.e. $\tilde{\beta}=-0.6$).}
\label{fig2}
\end{figure}

\clearpage
\begin{figure}
\begin{center}
\includegraphics[width=160mm]{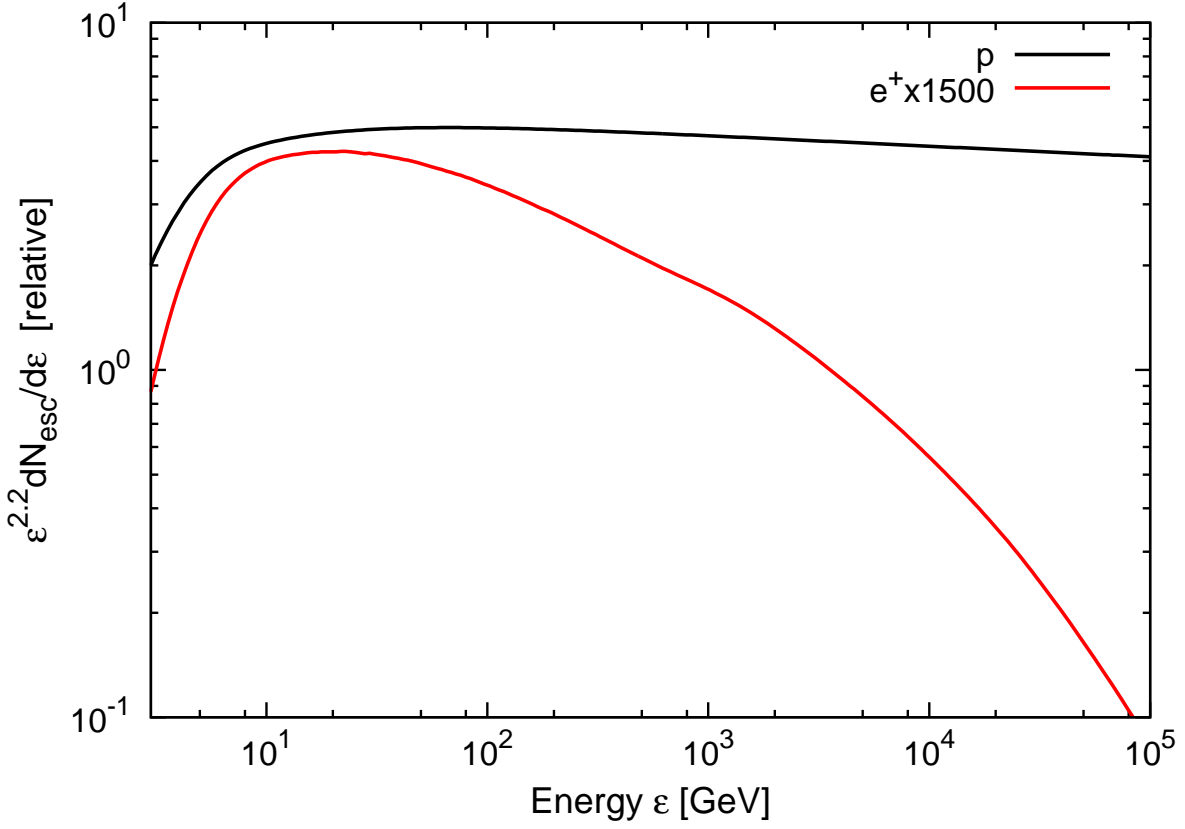}
\end{center}
\caption{Time-integrated energy spectra of protons (black) and positrons (red) escaping a SNR.  For convenience, the spectra are multiplied by $\varepsilon^{2.2}$, and as for the positron spectrum the overall factor of 1500 is multiplied.  The adopted parameters are the same as in Fig. 2.}
\label{fig3}
\end{figure}

\end{document}